# On the origin of Mount Etna eruptive cycles and Stromboli volcano paroxysms: implications for an alternative mechanism of volcanic eruption

Andrei Nechayev

Geographic Department, Moscow State University

*New mechanism of imbalance between magma column and fluid volume, accumulated in the magmatic system, is considered as a driving force of the volcanic eruption. Conditions of eruption based on this mechanism are used to explain main features of the volcanic activity (eruptive cycles and paroxysms) of the volcanoes Etna and Stromboli (Italy).*

**Introduction. Models of volcanic eruption. Mechanism of Gas-Liquid Imbalance.**

Despite the obvious progress in volcanology which occurred in the last 40 years (Parfitt and Wilson, 2008), the clear representation of the mechanism of volcanic eruption is still missing. In the area of basaltic volcanism two most popular models are: the Rise Speed Dependence (RSD)-model, where the acceleration and eruption occur due to magma defragmentation and expansion of gas bubbles; the Collapse Foam (CF)-model, where gas bubbles accumulated inside plumbing system attain some critical volume, rise with magma and erupt out from volcano (Parfitt, 2004). RSD-model allows extensive physical and mathematical treatment, since the formation of bubbles, reducing density and viscosity of magma melt, can be adequately considered in the thermo- and hydrodynamics of the process (Namiki and Manga, 2006). In the CF-model similar approach is missing. However, in (Del Bello et al., 2012) the possibility of bubble instability depending on its linear dimensions is demonstrated. In fact, CF-model is an analogue of geyser eruption model proposed in the early XIX century (Mackenzie, 1811). It is important that for RSD- model you need a primary upward movement of magma to bring at lower pressures new portions of gas-saturated melt. Further acceleration of magma (up to supersonic speeds) is explained by its defragmentation and decreasing of average density and viscosity due to gas bubbles. But where does this driving force of the primary magma ascent come from? This question usually remains unanswered. Due to lithostatic pressure of the crust? It is not, otherwise we would have already "bathed" in the magma. The average density of the continental crust is considerably less than the density of magma. The certainty in that the accelerating force is provided by gas bubbles disappears when you see the long-lived lava fountains of Hawaiian volcanoes in which the vertical jets of magma are slightly fragmented and therefore the melt density remains high. Bottle effect of "Coca-Cola" which supporters of RSD- model like to refer occurs only when you are opening a bottle sharply, that is when you quickly drop the pressure inside the liquid saturated with gas. Real magma rises to the surface slowly, its decompression occurs gradually (this, of course, not talking about explosive eruptions when lava vent is clogged by plug). As for the CF- model, it is unclear how a large volume gas bubble can push out magma column. It can float in the magma as a buoyant object or like a slug and explode near the surface, but that does not explain the lava fountains working for days and even months (Parfitt,2004).

A new physical mechanism of instability of the hydrostatic column of liquid contacting with a closed gas volume, recently discovered and described, offers an alternative approach to the main volcanological problem (Nechayev, 2012b). It was firstly proposed for explaining the geyser eruption (Nechayev, 2008). Recall the essence of this mechanism. If the liquid fills the tank with solid walls to the brim (vertical conduit) and has a contact with a closed gas volume at a depth $H$, their balance becomes unstable when the gas volume $V$ exceeds a critical value $V_{cr}$:

$$V > V_{cr} \equiv \gamma S(H + p_0 / \rho g) \qquad (1)$$

where $S$ is the conduit section in the contact area, $\gamma$ is the the adiabatic coefficient of the gas, $\rho$ is the liquid density, $p_0$ is the atmospheric pressure, $g$ is the acceleration of gravity.

Criterion (1) arises from the following simple considerations. Assume that the volume of the gas mass increased by a small amount $\Delta V$ due to the overpressure and penetrated into the conduit. In this case the same volume of liquid $\Delta V$ will be released from the conduit. $\Delta V$ is equal to $S\Delta z$, where $\Delta z$ is the reduction of the height of the liquid column, $S$ is the cross-section of the conduit. Corresponding decrease in the hydrostatic pressure of the liquid in contact point $\Delta p_l$ is equal: $\Delta p_l = \rho g \Delta z = -\rho g \Delta V / S$. The pressure in the gas volume also decreased adiabatically by the value $\Delta p_g$ which in accordance with the ideal gas law $p_g V^\gamma = A = const$ is equal to:

$$\Delta p_g = \frac{\partial p_g}{\partial V} \Delta V = -\frac{\gamma A}{V^{\gamma+1}} \Delta V \qquad (2)$$

Instability condition is: $|\Delta p_g| < |\Delta p_l|$. With (2) and the equality of the initial pressure of liquid and gas in the contact area we obtain the criterion (1).

Thus, during the gas expansion in the conduit (and removal of the appropriate volume of liquid) the pressure in the gas volume (when $V > V_{cr}$) remains greater than the hydrostatic pressure of liquid in the contact zone, so the liquid column will be erupted by the gas overpressure. It is important to stress that the critical condition (1) is applied to the area of contact only and is not dependent on the shape of the conduit above and the presence of the dilatations therein. It also does not depend on the viscosity of the liquid and gas, nor from their temperature. The described mechanism of imbalance was named in (Nechayev, 2012) as GP (gas-piston)-mechanism, but to distinguish it from gas-piston activity of Hawaiian volcanoes (Vergniolle and Mangan, 2000) we call it now the Gas-Liquid Imbalance - mechanism. GLI- mechanism is fundamental, it is easy to experimental verification and theoretical analysis. In (Nechayev, 2012a; Belousov et al, 2013) it was applied to the explanation of the physical nature of geyser eruption, in (Nechayev, 2012b) was extended to the case of a volcano eruption in presence of exogenous fluid (the fluid layer); a simple interpretation of the Caldera-forming eruptions has been demonstrated there. This mechanism is physically similar to the mechanism of bubble instability in the magma conduit described in (Del Bello et al., 2012). Uncontrolled expansion of the bubble when it reaches a certain critical length can be regarded as a special case of GLI- mechanism.

The aim of this work is to draw attention to the GLI-mechanism as probably a main link in the chain of eruption models, such as RSD, CF and other models. Here we will use the GLI-mechanism to explain the most powerful eruptions of basaltic volcanoes with endogenous (magmatic) fluid, taking as an example the well-known volcanoes Stromboli and Etna.

**Fluid overpressure in magma. Conditions of eruption.**

It must be recalled that in the case of geysers the role of gas substance in GLI-mechanism plays vapor accumulating under the vault of a subterranean chamber at water boiling. In the case of volcanoes water vapor or carbon dioxide in the supercritical state, probably, can serve as an ideal gas. They can form certain accumulations under the roof of magma chambers or within extensions and distortions of magma conduits.

Rising to the surface, magma "boils up": the "volatiles" dissolved in it are isolating in a separate gas phase dramatically increasing the pressure on magma and surrounding rocks. This pressure helps magma to move upwards. Mechanism for the gas bubbles generation resulting in the magma overpressure can be described as follows. The portion of fluid dissolved in magma $k$ decreases with depth $h$. When certain volume $V_m$ of magma with density $\rho_m$ moves upward at a distance $\Delta h$ the mass of fluid $\Delta M_f$ which has passed into the gas phase will be equal to:

$$\Delta M_f = V_m \rho_m \frac{\partial k}{\partial h} \Delta h. \qquad (3)$$

This mass will represent a lot of of micro-bubbles containing fluid with a density $\rho_f$ satisfying the ideal gas law and equal $P_l(h)/RT_m$, where $P_l(h)$ is the lithostatic pressure at a depth $h$; $T_m$ is the temperature of magma and $R$ is the gas constant. During ascent the lithostatic pressure decreases, and gas bubble will seek to expand, but having no possibility to increase its volume (due to the rigidity of the crust), the bubble will retain its original pressure exceeding the ambient pressure by an amount proportional $\rho_m g \Delta h$. This, in our view, encloses the physical nature of the internal overpressure in the magma. The total mass of the fluid isolated in the gas phase during magma ascent will be equal to the corresponding integral of the expression (3). For the first time fluid bubbles will be distributed uniformly across magma. But, as their density is always less than the magma density, they will float and accumulate in the appropriate places, for example, under the roof of the magma chamber (Fig. 1) .

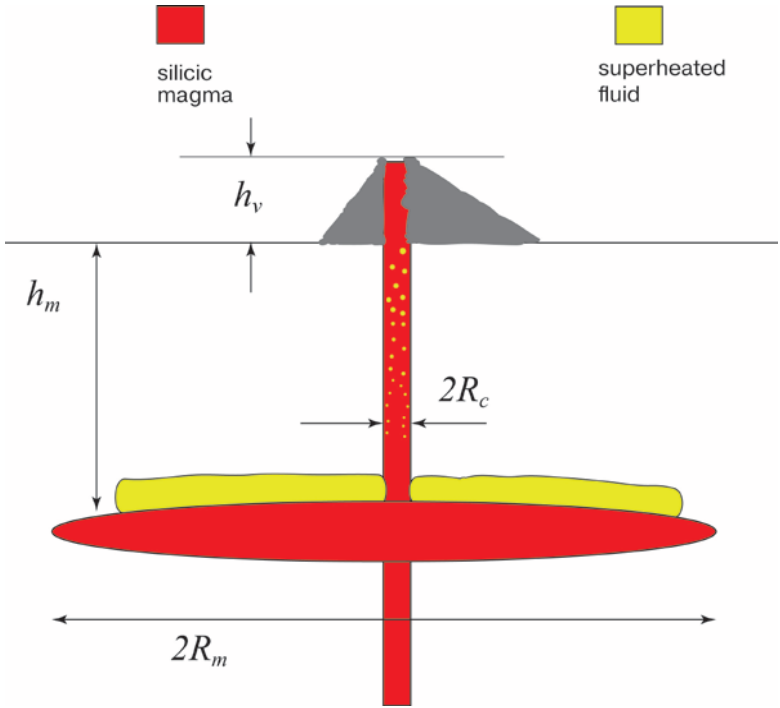

Fig.1 *Magma chamber with a "collar" of endogenous fluid. The section of the conduit: $S = \pi R_c^2$*

Increasing the pressure near the chamber roof, fluid will contribute to the formation of fractures, which enable magma to reach the surface. The maximum possible volume $V_f$ of the gaseous phase in the chamber will depend on the percentage of fluid in the melt $k(h)$ and the bedding depth of chamber $h_m$. With regard to (3) and the equality of pressures of magmatic column and fluid $\rho_m g(h_m + h_v) = \rho_f RT_m$ we obtain:

$$V_f = \frac{M_f}{\rho_f} = \frac{(k_0 - k_m)M_m}{\rho_f} = \frac{(k_0 - k_m)\rho_m V_m RT_m}{\rho_m g(h_m + h_v)} = \frac{(k_0 - k_m)V_m RT_m}{g(h_m + h_v)} \qquad (4)$$

where $M_m, V_m$ are the mass and the volume of the magma chamber, $(k_0 - k_m)$ - the portion of the fluid transformed in the gaseous phase at the level $h_m$ of magma chamber. In accordance with (1) the critical volume $V_{cr}$ for the case of Fig.1 is equal $\gamma(h_m + h_v)S$ as $p_0/\rho_m g \ll h_m$ for magma. Instability criterion $V_f > V_{cr}$ leads to the condition:

$$V_m > \frac{\gamma g S(h_m + h_v)^2}{(k_0 - k_m)RT_m} \qquad (5)$$

Thus, according to the GLI-mechanism, one of the necessary eruption conditions is directly related to the volume of the magma chamber, which must exceed a certain critical magnitude (5), depending on the depth of the chamber, its temperature and fluid content in the magma.

The second condition consists in the primary decreasing of magma column pressure that can occur when magma reaches the crater of the volcano (summit eruption) or finds a fracture on its slope (flank eruption). The further course of the eruption will depend on the volume of fluid as well as the parameters of the magmatic system. In any case the material (in whole or in part) located above the level $h_m$ will be erupted (Fig. 1).

**Etna and Stromboli. Eruptive cycles and "paroxysms."**
Etna is the largest and most active volcano in Europe, unsurpassed by its spectacular eruptions, as well as by the number of scientific papers devoted to it. However, despite Etna is available for large-scale observations and research, it arouses volcanologists around the world to have an inferiority complex keeping unsolved the mystery of its eruptive cycle. This cycle, which occupies the interval of 20-40 years, consists of series of summit eruptions, a series of flank eruptions and pauses in a few years, separating one cycle from another (Behncke and Neri, 2003). The final flank eruption of the series is always the most powerful.

Stromboli is not less known and popular than Etna. It impresses with the stable volcanic activity for centuries. His "routine" activity is small power explosions with gas and clots of lava emissions to a height of 50-200 m above the summit crater with an average frequency of 5-20 episodes per hour. More rare and more powerful explosions (the volume of erupted rocks about 100 m$^3$) occur average once a year, they are called "major explosion". Finally, about every 5 years the so-called "paroxysms" occur representing the powerful explosions with the eruptive column about one kilometer height and $10^4$-$10^6$ m$^3$ of erupted material (Andronico et al., 2008).

There are quite reliable geophysical data on the existence of two magma chambers under Etna at depths about 3-4 and 6-10 kilometers (Walter et al., 2005). Geo-chemical analysis of Stromboli eruptions also show that the sources of the erupted magma may be on two levels: deep (6-9 km) and shallow (2-4 km) (Aiuppa et al., 2011). We want to show that our GLI-mechanism allows to explain satisfactorily the features of the volcanic activity of Etna and Stromboli. Consider the diagram of a volcano plumbing system with two magma chambers (Fig. 2).

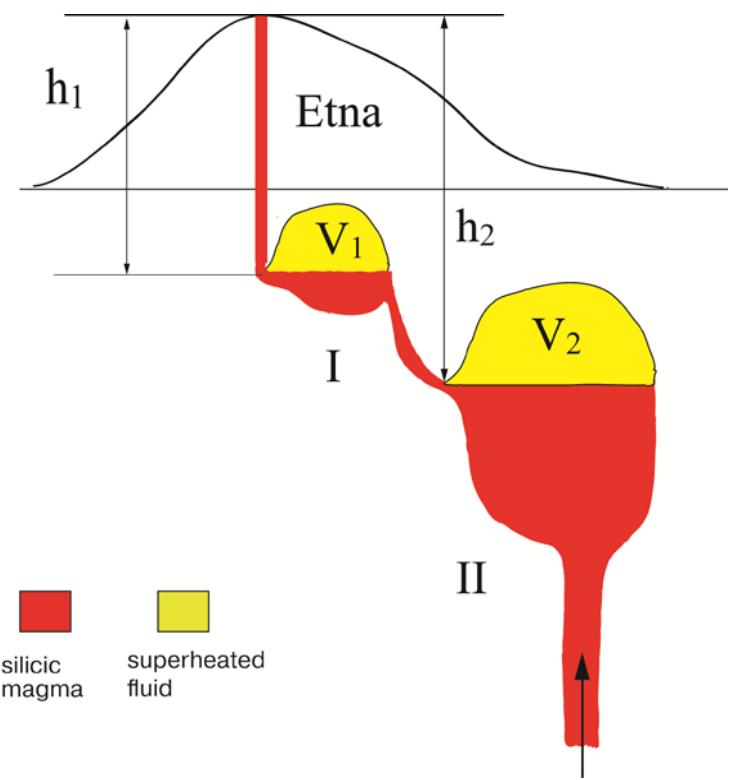

*Fig.2 Magma system with two communicating magma chambers: I - peripheral chamber supplying central conduit; II-intermediate chamber, representing the main source of magma*

Fresh magma enters the chamber II from deeper horizons, perhaps directly from the mantle. Periodicity of magma injection may be related to the features of magma intra- mantle convection. Filling the lower chamber II, magma releases a certain amount of fluid that occupies a volume $V_{f2}$ in the apical part of a chamber (Fig. 2) and originates there the overpressure under the influence of which magma is pushed to the upper chamber I and continues going to the crater of the volcano. Reducing the lithostatic pressure of the upper chamber by the amount of $\rho_m g(h_2 - h_1)$ the melt releases additional fluid volume $V_{f1}$ with corresponding overpressure.

If the volume of magma in the upper chamber exceeds a critical value (5), the volume of fluid will exceed $\gamma S h_1$ and the first requirement of the eruption will be satisfied. The second condition is the decreasing of the magma column pressure when magma reaches and overflows the volcano crater. Summit eruption should begin as soon as fluid penetrates into the conduit, it can discharge the total conduit contents above the upper chamber. As overpressure in the lower chamber persists, it must refill the magmatic system by the melt from the lower chamber, magma updates the upper chamber and rises to the top of the volcano. Thus, summit eruptions will be repeated until the overpressure in the volume $V_{f2}$ of lower chamber will be sufficent to fill the entire upper part of the magmatic system (Fig. 2), and the fluid will be enough to keep in the upper chamber its volume $V_{f1}$ exceeding the critical one. With each new summit eruption the volume of magma in the lower chamber will decrease (injection of magma from the mantle has not happened yet), respectively, will increase the volume of fluid $V_{f2}$ and decrease its pressure. The latter circumstance would cause the magma will not be able to reach the volcano's summit and will stand in the channel and find the weak spots on the side slope of the volcano for a breakthrough. Since the critical conditions for the upper chamber all being satisfied, and the magma column in the flank side gets the break to decrease the pressure, effusive eruptions continue on the slopes of the volcano. A series of flank eruptions could lead to the fact that the volume of fluid in the lower chamber, increasing after each eruption, exceeds a critical threshold, and at the time of its penetration in the conduit the new eruption will start, when all magmatic system above the lower chamber including its gas "collar" will be relieved. Naturally, this eruption must be the most powerful concerning the volume of erupted material and the last in the eruptive cycle as the fluid from the lower chamber will be exhausted. Pause in a series of eruptions must end with an updating of magma in the lower chamber. The above sequence of events is just recorded in numerous observations of the Mount Etna (Behncke and Neri, 2003).

Stromboli activity cycle may also begin with the filling of the lower (deeper) chamber by "fresh" magma that isolate a certain amount of fluid (mainly carbon dioxide), which accumulates at the roof of the chamber and occupies the initial volume $V_{f2}$ (Fig. 2). Assume that this volume is much less than the volume $V_{m2}$ of the magma chamber and has no contact with the magma conduit. Because of its overpressure the fluid starts to squeeze out magma from the chamber II and move it to the top of the volcano. In the area between two chambers the decompressed magma will release carbon dioxide and water vapor in the form of bubbles rising up and accumulating in the upper part of the chamber-trap I, which unlike the Etna supposedly has a volume less than the critical one. Another part of fluid released during the ascent of magma through the conduit rises directly to the crater of the volcano and "provide" the routine activity of Stromboli.

When fluid penetrates from the upper chamber into the conduit, the eruption does not occur, since the volume of fluid is always less than the critical value. Fluid penetrates into the conduit in the form of a bubble having a large linear dimensions and the pressure corresponding to its depth. Further motion of a bubble to the surface and its burst may follow the mechanism analyzed in [Del Bello et al., 2012], so the bubbles from the upper chamber may serve as the source of "major explosions". Thus, the cause of the weak and relatively strong explosions of Stromboli, as researchers believe (Andronico et al., 2008), is the rise of gas bubbles through the magmatic conduit and their destruction near the surface with the release of gas and magma clots. Explosive force must obviously depend on the value of the overpressure acquired by bubble at the corresponding depth. As for the paroxysms, their mechanism, in our opinion, has a different origin, although supporters of the "bubbles" theories explain such eruptions by non-standard sizes of bubble-slug (Del Bello et al., 2012; Jaupart and Vergniolle, 1989). The nontrivial fact characteristic for paroxysms was found by Calvari et al, 2011: it is the existence of a "cumulative" amount of lava, which must be erupted before paroxysmal events. Critical volume of lava was equal to $4 \cdot 10^6$ m$^3$. It was found for two paroxysms occurred April 5, 2003 and March 15, 2007 during the corresponding effusive eruptions (Calvari et al, 2011).

The process taking place in the lower chamber has, in our view, fundamentally different physics. The overpressure in the volume $V_{f2}$ must be sufficient for magma rising to the top of the volcano, for breaking through the volcano slope to organize the effusive eruption. Magma gradually leaves the lower chamber, so the volume of fluid $V_{f2}$ accordingly increases. The fluid pressure, of course, falls but it can be maintained by bubbles emerging from deep horizons of the magmatic system. Magma is squeezed from the chamber, the volume of gaseous fluid increases. Once fluid (most likely a carbon dioxide (Aiuppa et al., 2011)) will penetrate into the conduit, it may trigger the GLI- mechanism as the critical volume $\gamma S h_2$ can be surpassed. Further the fluid pressure in the contact zone will decrease more slowly than the pressure of the magma column. The pressure drop firstly will be minimum but over time it should increase. Perhaps in this initial period of magma acceleration the abnormal $CO_2$ emissions from the crater must be observed (Aiuppa et al., 2011) as carbon dioxide under overpressure intensively enters the conduit from the lower chamber. Hydrodynamics of the eruption process having a nonlinear nature may well be catastrophic because at this stage it should join the defragmentation of magma melt. Accordingly all material contained in the conduit and all its dilatations above the level $h_2$ will be erupted.

In accordance with the GLI-mechanism the critical volume of erupted lava that precedes the "paroxysm" must be equal to the volume increase, which obtain the fluid in the lower chamber during effusive eruptions before penetrating the magma conduit. From this moment the paroxysmal eruption of the melt located in the conduit above $h_2$ depth must begin.

**Concluding remarks.**
What could be the real magnitude of the critical volume? Let us evaluate it. For the lower chamber of the supposed magmatic system of Stromboli (Fig. 2) it should be equal $\gamma S h_2$. Taking the conduit diameter as 4 m (at April 5, 2003 paroxysm (Calvari et al., 2005) a video camera fixed the glowing object with a diameter of 4 m above the crater), the depth as 6 km and carbon dioxide as the fluid ($\gamma = 1,3$ ), we obtain the critical volume equal to $10^5$ m$^3$. If we assume that the fluid is distributed uniformly along the roof of the magma chamber (Fig. 1), then for a typical chamber radius of 1 km we obtain the average thickness of the fluid "collar" just only 0.03 m! It is clear that the fluid under the roof will accumulate non-uniformly, but from the above calculation it is evident that the geophysical methods can't fix such fluid accumulation even it exceed this amount hundreds of times.

Critical volume "works" only when the fluid penetrates the magma conduit. The greater the excess of the actual volume of fluid above the critical one, the more intense eruption develops. Amount of magma erupted before the paroxysm of Stromboli in 2003 was equal to $4 \cdot 10^6 m^3$, and it gives a minimum evaluation of the expansion volume of the fluid in the lower chamber. So this volume could exceed the critical value of at least an order of magnitude. The entire volume of erupted magma during the paroxysm was 84 000 $m^3$ (Calvari et al., 2005), which corresponds to the "idealized" channel depth of 6 km and a diameter of 4 m.

Effusive eruption of Stromboli before "paroxysm" represents an analogue of the classical overflow phase before geyser eruption, when the water vapor accumulating under the arch of the underground chamber, quietly pushes the water out of the conduit as long as it begins to penetrate into the conduit (Nechayev, 2012a). In the case of Stromboli the similar "trigger" role may play the carbon dioxide accumulated under the roof of the lower magma chamber (Fig. 2). Significant quantity of $CO_2$ captured in the "traps" of Stromboli magmatic system were reported in (Aiuppa et al., 2011): it amounted to 11,000 tons which corresponds to the gas volume of $10^4$ $m^3$ at a depth of 6 km. This volume is ten times less than the critical volume for the lower chamber, so it could not trigger the paroxysm but it could manage two "major explosion" from chamber I which took place on 25 and 30 June 2010 (Aiuppa et al., 2011). About the volumes of carbon dioxide accumulated by magma system before Strombolian "paroxysms" nothing is known.

Thus, the popular CF and RSD-models of basaltic volcanic eruptions (Parfitt, 2004) probably represent the united chain with the above described GLI-mechanism: the formation of a big "bubble" from the gas foam in CF-model is a stage of the critical volume attainment, the collapse is the onset of instability and RSD-model with appropriate magma defragmentation joins GLI-mechanism at the time of dynamical motion and acceleration of the magma column.